\documentclass[10pt,a4paper]{article}
\usepackage[english]{babel}
\usepackage[utf8]{inputenc}
\usepackage{amsfonts,amsbsy,bm,euscript,mathrsfs}
\usepackage{amssymb,stmaryrd,faktor}
\usepackage[tbtags]{amsmath}
\usepackage{bbm}
\usepackage{float}
\usepackage{graphicx}
\usepackage[title,titletoc]{appendix}
\usepackage[bookmarks=true,colorlinks=true,linkcolor=blue,citecolor=blue,urlcolor=blue,bookmarksnumbered]{hyperref}
\usepackage{multicol}

\usepackage{dsfont}
\usepackage{lmodern}
\usepackage{mathrsfs}
\usepackage{mathtools}
\usepackage{bbm}
\usepackage{braket}
\usepackage{slashed}

\usepackage{slashbox} 

\usepackage{arydshln}

\usepackage{tikz-cd}

\numberwithin{equation}{section}

\makeatletter
\renewcommand\section{\@startsection {section}{1}{\z@}
{-3.5ex \@plus -1ex \@minus -.2ex}
{2.3ex \@plus.2ex}
{\normalfont\Large\bfseries}}
\renewcommand\subsection{\@startsection{subsection}{2}{\z@}
{-3.25ex\@plus -1ex \@minus -.2ex}
{1.5ex \@plus.2ex}
{\normalfont\large\bfseries}}
\makeatother

\newcommand{\bea}{\begin{eqnarray}}
\newcommand{\eea}{\end{eqnarray}}

\allowdisplaybreaks 

\usepackage{amsthm}

\usepackage{blkarray}

\begin{document}

\thispagestyle{empty}
\begin{flushright}\footnotesize\ttfamily
DMUS-MP-22/07
\end{flushright}
\vspace{2em}

\begin{center}

{\Large\bf \vspace{0.2cm}
{\color{black} \large Jordan blocks and the Bethe ansatz II: The eclectic spin chain beyond $K=1$}} 
\vspace{1.5cm}

\textrm{Juan Miguel Nieto García\footnote{\texttt{j.nietogarcia@surrey.ac.uk}}}

\vspace{2em}

\vspace{1em}
\begingroup\itshape
Department of Mathematics, University of Surrey, Guildford, GU2 7XH, UK
\par\endgroup

\end{center}

\vspace{2em}

\begin{abstract}\noindent 
We continue the classification of the Jordan chains of the eclectic three state spin chain that we started in our previous article. Following the same steps, we construct the generalised eigenvectors of this spin chain by computing the strongly twisted limit of linear combinations of eigenvectors of a twisted XXX $SU(3)$ spin chain. We show that this classification problem can be mapped to the computation of the number of positive integer solutions of a system of linear equations.
\end{abstract}

\newpage

\overfullrule=0pt
\parskip=2pt
\parindent=12pt
\headheight=0.0in \headsep=0.0in \topmargin=0.0in \oddsidemargin=0in

\vspace{-3cm}
\thispagestyle{empty}
\vspace{-1cm}

\tableofcontents

\setcounter{footnote}{0}

\section{Introduction}

Non-Hermitian systems have a wide range of applications in physics, ranging from optics to critical phenomena, even appearing in the context of transport phenomena in biological systems, see \cite{Heiss:2012dx,Ashida:2020dkc} and references therein for particular instances. In contrast, this kind of systems are not as widely studied in the context of quantum mechanics due to the Dirac-von Neumann axiom regarding the Hermiticity of the Hamiltonian. Nevertheless, it has been shown that this axiom can be related to the existence of an anti-linear operator that commutes with the Hamiltonian (for example, $\mathcal{PT}$ symmetry) \cite{Bender:1998ke,Bender:2007nj,Alexandre:2015kra,PT}. Non-Hermitian systems are also important in the context of Conformal Field Theory, as correlation functions with logarithmic singularities can only arise from a Virasoro operator $L_0$ with non-trivial Jordan cells (which cannot happen if $L_0$ is Hermitian) \cite{Gurarie:1993xq}. Furthermore, it has been shown that non-Hermitian Conformal Field Theories do not have to satisfy Zamolodchikov's c-theorem (one example is the non-Hermitian Sine-Gordon Model \cite{Fendley:1993wq,Ashida:2016}).

Although there has always been an interest in the topic of non-Hermitian system in the field of integrability, this interest has grown in recent times driven by the fascinating properties of a particular deformation of $AdS_5 \times S^5$ called \emph{fishnet theory}, proposed in \cite{Gurdogan:2015csr}. Strongly twisted theories, which contain the fishnet theory, are obtained from the $\gamma_i$ deformation of $AdS_5 \times S^5$ \cite{Lunin:2005jy,Frolov:2005dj,Fokken:2013aea,Fokken:2013mza} by considering the infinite imaginary twist and vanishing coupling regime while keeping their product constant, $e^{-i\gamma_j/2} \rightarrow \infty$, $g\rightarrow 0$, $e^{-i\gamma_j/2} g=\xi_j$. The most general of these strongly twisted theories has the following interaction Lagrangian (up to relabelling of the fields)
\begin{align}
 \mathcal{L}_{\text{int}} =&-i N \text{Tr} \left[ \sqrt{\xi_2 \xi_3} (\psi^3 \phi^1 \psi^2 +\bar{\psi}_3 \phi_1^\dagger \bar{\psi}_2 ) + \sqrt{\xi_3 \xi_1} (\psi^1 \phi^2 \psi^3 +\bar{\psi}_1 \phi_2^\dagger \bar{\psi}_3 ) + \sqrt{\xi_1 \xi_2} (\psi^2 \phi^3 \psi^1 +\bar{\psi}_2 \phi_3^\dagger \bar{\psi}_1 ) \right] \notag \\
 &- N \text{Tr} \left[ (\xi_3)^2 \phi_1^\dagger \phi_2^\dagger \phi^1 \phi^2 + (\xi_1)^2 \phi_2^\dagger \phi_3^\dagger \phi^2 \phi^3 + (\xi_2)^2 \phi_3^\dagger \phi_1^\dagger \phi^3 \phi^1 \right]   \ , \label{interactionstrongtwist}
\end{align}
where $\phi^i$ are complex bosonic fields and $\psi^j$ are fermionic fields. The gauge fields and the fourth fermion decouple in this limit. The fishnet theory is obtained by turning off all but one of the deformation parameters $\xi_i$.

Similarly to $\mathcal{N}=4$ SYM theory, we can express the action of the dilatation operator on single trace operators at one-loop in terms of the nearest-neighbour Hamiltonian of an effective spin chain \cite{Ipsen:2018fmu}. Direct inspection of the Lagrangian~(\ref{interactionstrongtwist}) is enough to realise that these strongly twisted theories are non-Hermitian, which implies that the dilatation operator associated with the conformal symmetry of the theory is no longer Hermitian for general values of the deformation parameters. It has been observed that not only do the eigenvalues associated with the dilatation operator take complex values, but that it also becomes non-diagonalisable. The effective spin chain that describes the action of the one-loop dilatation operator on single trace operators made of scalars was first studied in \cite{Ipsen:2018fmu,StaudacherAhn}. To that end, the authors solved the Bethe equations for finite values of the twist parameter $q_i$ and computed the large twist limit. Although they were able to find the correct number of Bethe roots, they were not able to find all the Bethe vectors. The reason for it was that several Bethe vectors have the same limit at strong twist. This is an effect particular to non-Hermitian matrices called \emph{coalescence}, and it can be understood as a version of degeneracy for eigenvectors. This spin chain has been further studied in \cite{Ahn:2021emp}, where the full structure of the Jordan chains was untangled (up to some subtleties that we will discuss later) using combinatorial arguments.

In our previous article \cite{firstpart} we also addressed this problem from the same perspective as \cite{StaudacherAhn} by considering the limiting procedure in more detail. In said article, we proved that it is necessary to consider the limit of a linear combination of eigenvectors of the theory at finite twist if we want to reconstruct a generalised eigenvector of the eclectic spin chain. In fact, we showed that for any diagonalisable matrix that depends on a complex parameter, we need to consider the limit at an exceptional point (i.e. a value of the parameter where the matrix becomes non-diagonalisable) of at least $m$ eigevectors if we want to compute a generalised eigevector of rank $m$ of said matrix at this exceptional point. After some tedious computations, we were able to untangle the Jordan chain structure for the subsector of operators with exactly one scalar $\phi^3$. The results matched those found nearly simultaneously in \cite{Ahn:2021emp}. As the two methods to obtain information about the Jordan chains are completely different, the matching results were a welcomed conformation.

In this article, we plan to extend the results of \cite{firstpart} beyond the subsector containing exactly one scalar $\phi^3$. Although the structure of the Jordan chains in these subsectors has already been studied in \cite{Ahn:2021emp}, the method we will present here to compute them is different, so it is worth checking if both results are consistent.

The outline of this article is as follows.  In section 2 we review the connection between the strong twist limit of $\gamma_i$-deformed $\mathcal{N}=4$ SYM and the eclectic spin chains introduced in \cite{Ipsen:2018fmu}. In section 3 we summarise the results from \cite{firstpart} regarding how to construct the generalised eigenvectors of a matrix at an exceptional point. In section 4 we describe how to construct the Nested Coordinate Bethe Ansatz for the twisted $SU(3)$ spin chain for any number of excitations. In section 5 we apply the procedure from section 3 to the eigenstates of the finite twisted spin chain computed in section 4 to find the generalised eigenvectors of the eclectic spin chain. We show that we can map the problem of computing how many generalised eigenvectors we have to the number of solutions of a system of linear Diophantine equations. In section 6 we explain in detail how to compute the number of solutions of those systems of equations. Section 7 closes this article with a summary and some final remarks.

\section{Twisted spin chain and eclectic spin chain} \label{themodel}

A very well-known procedure to compute the conformal dimension of operators in $\mathcal{N}=4$ SYM is to construct an effective spin chain whose Hamiltonian acts in the same fashion as the dilatation operator acts on single-trace operators. A detailed review of this method can be found in \cite{Minahan:2010js}. This procedure can be extended to the $\gamma_i$-deformation, and thus to the strong twist deformation of $\mathcal{N}=4$ SYM we are interested in. The details of this construction can be found in \cite{Fokken:2013aea,Fokken:2013mza,Ipsen:2018fmu,StaudacherAhn}. Here we will follow the expressions from the last reference.

If we consider only operators constructed with three different kinds of scalar fields, the effective spin chain associated to the dilatation operator of the $\gamma_i$-deformation of $\mathcal{N}=4$ SYM is an integrable deformation of the $SU(3)$ XXX spin chain that appear in regular $\mathcal{N}=4$ SYM. This integrable deformation is called \emph{twisting} and, for the particular case of $SU(3)$, we can introduce a total of 3 different twist parameters, $q_i$. Similarly to the untwisted case, the dilatation operator acting on the trace of $L$ scalar fields can be written as
\begin{equation}\label{eq:twistedXXX}
	\mathcal{D}=\mathcal{D}_0 + g^2 \mathbf{\tilde{H}}_{q_1 , q_2 , q_3}+\mathcal{O} (g^4) =\mathcal{D}_0 + g^2 \left[\sum_{l=1}^L \tilde{\mathbb{P}}^{l,l+1} \right]+\mathcal{O} (g^4) \ ,
\end{equation}
where $\mathcal{D}_0$ is the bare dimension of the operator and $\tilde{\mathbb{P}}^{a,b}$ is a \emph{twisted permutation operator}. If we identify the scalar field $\phi^i$ with the spin state $|i\rangle$, this operator $\tilde{\mathbb{P}}^{a,b}$ acts non-trivially on sites $a$ and $b$ as follows
\begin{align}\label{eq:twistedperm}
\tilde{\mathbb{P}}\, |11\rangle &= |11\rangle \ , & \tilde{\mathbb{P}}\,|22\rangle &= |22\rangle \ , & \tilde{\mathbb{P}}\,|33\rangle &= |33\rangle \ , \\ \nonumber
\tilde{\mathbb{P}}\,|12\rangle &=\frac{1}{q_3}\, |21\rangle \ , & \tilde{\mathbb{P}}\,|23\rangle &=\frac{1}{q_1}\, |32\rangle \ , & \tilde{\mathbb{P}}\,|31\rangle &=\frac{1}{q_2}\, |13\rangle \ , \\ \nonumber
\tilde{\mathbb{P}}\,|21\rangle &=q_3\, |12\rangle \ , & \tilde{\mathbb{P}}\,|32\rangle &=q_1\, |23\rangle \ , & \tilde{\mathbb{P}}\, |13\rangle &=q_2\, |31\rangle \ .
\end{align}
In addition, as we want to consider only single-trace operators, we will work with closed spin chains with the periodic identification $L+1\equiv 1$. Although this Hamiltonian is Hermitian only if the twist parameters are complex phases, it is diagonalisable for generic values of said parameter.

The eclectic spin chain we are interested in is associated to the strong twist and weak coupling regime of the $\gamma_i$-deformation, so the Hamiltonian of the eclectic spin chain can be obtained as the following limit
\begin{equation}
	\mathbf{\hat{H}}_{\xi_1 , \xi_2 , \xi_3}=\lim_{\epsilon \rightarrow 0} \epsilon \mathbf{\tilde{H}}_{(\frac{\xi_1}{\epsilon},\frac{\xi_2}{\epsilon},\frac{\xi_3}{\epsilon})} =\sum_{l=1}^L  \hat{\mathbb{P}}^{l,l+1}\ ,
\end{equation}
where $\hat{\mathbb{P}}^{a,b}$ is an operator that acts non-trivially on sites $a$ and $b$ as follows
\begin{align}\label{eclecticperm}
\hat{\mathbb{P}}\,|21\rangle &=\xi_3\, |12\rangle \ , & \hat{\mathbb{P}}\,|32\rangle &=\xi_1\, |23\rangle \ , & \hat{\mathbb{P}}\, |13\rangle &=\xi_2\, |31\rangle \ ,
\end{align}
while the remaining matrix elements are zero.

As said above, the Hamiltonian $\mathbf{\tilde{H}}$ is an integrable deformation of an XXX spin chain, so we can apply the usual Bethe ansatz to construct its eigenvalues and eigenvectors. Thus, rather than computing the eigenvalues and eigenvectors of the eclectic spin Hamiltonian $\mathbf{\hat{H}}$ brute force thought the usual procedure, our approach will be to compute them as the strong twist limit of the eigenvalues and eigenvectors of $\mathbf{\tilde{H}}$. Sadly, despite the simple relation between the two Hamiltonians, the relation between the two spectra is not that simple. If we straightforwardly compute the limit of the eigenvalues, we will find no problem, but if we do the same for the eigenvectors, we will soon realise that we are not able to span the complete Hilbert space. This happens because the strongly twisted limit is an exceptional point (that is, the Hamiltonian becomes non-diagonalisable in this limit), and eigenvectors will start to coalesce as we approach it. The driving force behind this coalescence is the fact that the limit of an eigenvector has to be an eigenvector, together with the obvious observation that a non-diagonalisable matrix has fewer linearly independent eigenvectors than a diagonalisable matrix of the same dimension by definition.

If we explicitly do the calculation, we find that the finitely twisted vectors coalesce into states of the form
\begin{equation}
	|\psi (p) \rangle=\sum_{l=1}^L e^{2\pi p l i /L} U^l | 1, \dots, 1, 2,\dots , 2, 3, \dots, 3 \rangle \label{lockedstate}
\end{equation}
where $L$ is the total number of sites and $U$ is the translation operator that moves every site of the spin chain one position to the left, i.e., $U |n_1, \dots ,n_{L-1}, n_L \rangle=|n_2, \dots ,n_L, n_1 \rangle$. Due to their form, where the excitations cannot move from their relative places, these states have been called \emph{locked states}. Although we can check that they are eigenvectors of the Hamiltonian $\mathbf{\hat{H}}_{\xi_1 , \xi_2 , \xi_3}$, we can also check that these cannot be all the eigenvectors of this Hamiltonian. Obviously, this naïve computation gives us zero information about generalised eigenvectors, but the real problem is that it does not provide us with the full set of eigenvectors. We address with this topic in the following section.

In the remaining parts of this article, we will consider the single trace operator build solely of $\phi^1$ fields as the vacuum state of our effective spin chain. This means that a scalar $\phi^2$ behaves as a right-moving excitation, while a scalar $\phi^3$ behaves as a left-moving excitation. We will denote the total number of excitations, i.e. the number of $\phi^2$ and $\phi^3$ fields in the operator, by $M$, and the total number of $\phi^3$ fields in the operator by $K$. Without any loss of generality, we can assume that $K\leq M-K \leq L-M$.

\section{Generalised eigenvectors from the limit of diagonalisable matrices} \label{themethod}

In this section, we plan to review the method to construct generalised eigenvectors presented in our previous article \cite{firstpart}, which extends the recipe suggested in \cite{Gainutdinov:2016pxy} to Jordan blocks of size larger than two.

First, let us recapitulate some concepts regarding diagonalisability of matrices. Given a matrix $M$, we say that $\lambda_i$ is an eigenvalue of $M$ with \emph{algebraic multiplicity} $n_i$ if it is a zero of degree $n_i$ of the characteristic polynomial of $M$, that is, if
\begin{equation}
	\det (M-\lambda \mathbb{I} ) \propto (\lambda - \lambda_i)^{n_i} \ ,
\end{equation}
where $\mathbb{I}$ is the identity matrix. We say that $v_i$ is an eigenvector of $M$ associated with the eigenvalue $\lambda_i$ if
\begin{equation}
	M v_i=\lambda_i v_i \ .
\end{equation}
The total number of linearly independent vectors that fulfil this equation, i.e. the dimension of $\text{Ker}(M-\lambda_i I)$, is called \emph{geometric multiplicity} of $\lambda_i$. It is easy to check that the geometric multiplicity is either equal or smaller than the algebraic multiplicity.

If the geometric multiplicity of each of the eigenvalues is equal to its corresponding algebraic multiplicity, there exists a similarity transformation that makes the matrix diagonal, and we say that the matrix is \emph{diagonalisable}. Instead, if the geometric multiplicity of at least one eigenvalue is smaller than its algebraic multiplicity, we say that the matrix is \emph{non-diagonalisable} or \emph{defective}, as there does not exist a similarity transformation that makes the matrix diagonal. The next best thing we can do is to write a similarity transformation that changes a given square matrix into a block diagonal matrix of the form
\begin{equation}
	M=\begin{pmatrix}
	J_1 & 0 & 0 &   \dots \\
	0 & J_2 & 0 &  \dots \\
	0 & 0 & J_3  & \dots \\
	\vdots & \vdots & \vdots & \ddots
	\end{pmatrix} \ ,  \text{ which each block being of the form } J_i=\begin{pmatrix}
	\lambda_i & 1 & 0 & 0 &  \dots \\
	0 & \lambda_i & 1 & 0 & \dots \\
	0 & 0 & \lambda_i & 1 & \dots \\
	0 & 0 & 0 & \lambda_i & \dots \\
	\vdots & \vdots & \vdots & \vdots & \ddots
	\end{pmatrix} \ .
\end{equation}
This matrix is called the \emph{Jordan normal form} of $M$, and each block is called \emph{Jordan block} or \emph{Jordan cell}.

In order to find this similarity transformation, we have to introduce the  concept of \emph{generalised eigenvectors} and \emph{Jordan chains}. Given a defective eigenvalue $\lambda_i$ and an eigenvector $v_{i,\alpha}^{(1)}$ associated with it, we define the \emph{generalised eigenvector of rank $n$} as the vector fulfilling
\begin{equation}
	(M-\lambda_i \mathbb{I} ) v_{i,\alpha}^{(n)}=v_{i,\alpha}^{(n-1)} \ , \label{geneigenvdefinition}
\end{equation}
where the index $\alpha$ labels the possible geometric multiplicity of the eigenvalue $\lambda_i$. It is trivial to prove that these generalised eigenvectors also fulfil $(M-\lambda_i \mathbb{I} )^n v_{i,\alpha}^{(n)}=0$. The set of these interconnected generalised eigenvectors are said to form a (right) Jordan chain.  Notice that the generalised eigenvectors that form a Jordan chain are linearly independent but not necessarily orthogonal. A Jordan chain is said to have length $k$ if it contains $k$ generalised eigenvectors. For simplicity, we will usually drop the $(1)$ superindex for eigenvectors.

Let us consider now a matrix $M(\epsilon )$ that depends on a complex parameter and assume that it is diagonalisable for almost all values of $\epsilon$. The values of $\epsilon$ at which the matrix is not diagonalisable are called \emph{exceptional points}.

For simplicity, let assume that $\epsilon=0$ is an exceptional point of our matrix. We will denote the generalised eigenvector of rank $n$ of $M(0)$ associated to the eigenvalue $\mu_i$ as $u_i^{(n)}$. Outside its exceptional points, we will denote the eigenvalues and eigenvectors of $M(\epsilon )$ by $\lambda_i$ and $v_i$ respectively. Although they depend explicitly on $\epsilon$, we will not explicitly write that dependence most of the time, as it will clutter our expressions.

If we compute the limit of the eigenvector equation at the exceptional point, we find that
\begin{equation}
	0=\lim_{\epsilon\rightarrow 0} \left[(M(\epsilon)-\lambda_i \mathbb{I}) v_i\right]= \left[\lim_{\epsilon\rightarrow 0}(M(\epsilon)-\lambda_i \mathbb{I})\right]  \left[ \lim_{\epsilon\rightarrow 0}v_i\right]= \left[M(0)- \mathbb{I} \mu_i\right]  \left[ \lim_{\epsilon\rightarrow 0}v_i\right] \ . \label{important}
\end{equation}
This is exactly the same as the eigenvector equation for $M(0)$, thus the vector space spanned by the limit of the eigenvectors of $M(\epsilon)$ has to be a subspace of the vector space spanned by the eigenvectors of $M(0)$. We should stress that the dimensions of these two vectors spaces are not guaranteed to be the same unless the eigenvector $\mu_i$ only has a single Jordan chain associated to it.

Let us consider two eigenvectors $v_1$ and $v_2$ such that their eigenvalues do not necessarily coincide, but their limit does, $\lim_{\epsilon\rightarrow 0} \lambda_1 = \lim_{\epsilon\rightarrow 0} \lambda_2 = \mu_i$. Although it may look like this requirement is very restrictive, it is actually not. If two eigenvalues end up in the same Jordan block, they have to satisfy the previous requirement by definition of Jordan block. As we are assuming that $M(0)$ is non-diagonalisable, there has to be at least two eigenvectors that fulfils this requirement. Now, as both $v_1$ and $v_2$ are eigenvectors of $M(\epsilon)$, it is immediate to check that a linear combination of them fulfils
\begin{equation}
	[M(\epsilon) - \lambda_1 \mathbb{I} ][M(\epsilon) - \lambda_2 \mathbb{I} ] (\alpha_1 \, v_1 + \alpha_2 \, v_2)=0 \ .
\end{equation}
If we compute the limit of this equation at the exceptional point, we will find
\begin{equation}
	\lim_{\epsilon\rightarrow 0} \left[M(\epsilon )- \mathbb{I} \lambda_1\right] \left[M(\epsilon )- \mathbb{I}\lambda_2\right]  \left[ \lim_{\epsilon\rightarrow 0} (\alpha_1 \, v_1 + \alpha_2 \, v_2) \right]=\left[M(0)- \mathbb{I} \mu_i \right]^2  \left[ \lim_{\epsilon\rightarrow 0} (\alpha_1 \, v_1 + \alpha_2 \, v_2) \right]=0 \ , \label{GEEquation}
\end{equation}
meaning that this linear combination contains information about the generalised eigenvectors of rank $2$. Similar relations hold if we consider the linear combinations of more eigenvectors of $M(\epsilon)$. 

Although a linear combination of $k$ eigenvectors contains information regarding the generalised eigenvectors up to rank $k$, it is not immediately obtainable because information about all of them is encoded therein. We found that the appropriate method to extract this information for the case of $\mu_i$ being associated to only one Jordan chain is to consider the limit of the following linear combinations
\begin{equation}
	w_{ij}^{(n)}=\frac{ w^{(n-1)}_{ji} - \beta^{(n-1)}_{kj} w^{(n-1)}_{ki}}{|w^{(n-1)}_{ji} - \beta^{(n-1)}_{kj} w^{(n-1)}_{ki}|} \qquad \text{with} \qquad \beta^{(n-1)}_{kj}= (w^{(n-1)}_{ji})^\dagger \cdot w^{(n-1)}_{ki}  \quad \text{and} \quad w^{(0)}_{ij}=v_i\ , \label{recipe}
\end{equation}
where $|v|$ is the usual norm of complex vectors. If $\mu_i$ is associated to more than one Jordan chain, the procedure is similar, substituting $w^{(n-1)}$ by an appropriate linear combination of all the $w^{(n-1)}$ that we found at the $(n-1)$-th step and that give rise to linearly independent vectors at the exceptional point, while the $\beta^{(n-1)}$ coefficients are fixed by demanding orthogonality between these $w^{(n-1)}$ and the vector $w^{(n)}$ we are constructing.

If the eigenvalue $\mu_i$ is associated to only one Jordan chain, there is a one-to-one correspondence between the limit of the vectors $w^{(n)}$ and the generalised eigenvector of rank $n+1$ of $M(0)$. However, if $\mu_i$ is associated to two or more Jordan chains, this identification is not straightforward, and the different chains can start to mix. This is very clearly seen in the following four matrices
	\begin{align}
	&\begin{pmatrix}
	0 & 1 & \epsilon^2 \\
	\epsilon^2 & \epsilon^4 & \epsilon^2 \\
	0 & 0 & \epsilon^6
	\end{pmatrix} & &\begin{pmatrix}
	0 & 1 & \epsilon^2 \\
	\epsilon^2 & \epsilon^4 & \epsilon \\
	0 & 0 & \epsilon^6
	\end{pmatrix} & &\begin{pmatrix}
	0 & 1 & \epsilon^2 \\
	\epsilon^5 & 0 & 0 \\
	\epsilon & 0 & 0
	\end{pmatrix} & \begin{pmatrix}
	0 & 1 & \epsilon^2 \\
	0 & \epsilon^4 & \epsilon^2 \\
	0 & 0 & \epsilon^6
	\end{pmatrix} \ . \label{EX1}
\end{align}
All four matrices become the same at $\epsilon=0$, a matrix with two Jordan chains associated to the eigenvalue $\mu_i=0$. However, the set of $w^{(n)}$ vectors that we have is different in each case. The structure can be summarised in the following four diagrams, where every column of vectors corresponds to the limit of $w^{(n)}$ ordered in increasing values of $n$ from left to right,
	\begin{align*}
	&\begin{tikzcd}[ampersand replacement=\&]
   \hat{u}_1 \arrow[r] \& \hat{u}_2  \\
   \hat{u}_3 \arrow[ru]
\end{tikzcd} & &\begin{tikzcd}[ampersand replacement=\&]
   \hat{u}_1 \arrow[r] \arrow[rd] \& \hat{u}_2 \\
   \& \hat{u}_3
\end{tikzcd} & &\begin{tikzcd}[ampersand replacement=\&]
   \hat{u}_3 \arrow[r] \arrow[r] \& \hat{u}_1 \arrow[r]  \& \hat{u}_2
\end{tikzcd} & &\begin{tikzcd}[ampersand replacement=\&]
   \hat{u}_1 \arrow[r] \arrow[r] \& \hat{u}_2 \arrow[r]  \& \hat{u}_3
\end{tikzcd}
\end{align*}
where $\hat{u}_1=(1,0,0)$, $\hat{u}_2=(0,1,0)$, and $\hat{u}_3=(0,0,1)$. In the first two cases, we can distinguish that there are two Jordan chains (although in the second one, the second eigenvector appears together with the generalised eigenvector of rank $2$). In the latter two cases, one Jordan chain appears after the other, so our recipe cannot distinguish between them. We call this effect \emph{chain mixing}, and it has been discussed in detail both in appendix A of \cite{Ahn:2021emp} and section 5.2 of \cite{firstpart}. In particular, a method to deal with it was proposed in the latter reference.

In \cite{Ahn:2021emp} it was conjectured that chain mixing does not spoil the Jordan chain structure obtained by naïve computation for the eclectic spin chain. Although there is no formal proof of this statement, there is numerical evidence up to $L=30$ and $M=6$ and a formal proof for some specific Jordan chains. In the analysis of the Jordan chains of the eclectic spin chain we will present in the following section we will assume this conjecture to be true and we will not concern ourselves with the issue of chain mixing.

\section{The Coordinate Bethe Ansatz for Twisted Spin Chains}

In order to apply the method from the previous section to the eclectic spin chain, our first goal will be  to construct the eigenvectors of the twisted Hamiltonian (\ref{eq:twistedXXX}) for general values of the twist parameters $q_i$. The most straightforward method is a slightly modified version of the Nested Coordinate Bethe Ansatz (NCBA). The original method is described in section II.O of \cite{Sutherland}, as well as in section 2 of \cite{Beisert:2005fw}.\footnote{Another version of the NCBA using Zamolodchikov-Faddeev operators can be found in \cite{MariusZF}.}

First, we will present the case of $K=1$, which we have already discussed extensively in our first article on this topic \cite{firstpart}. After that, we will construct the $K=2$ case building on those results. Finally, we will show how to generalise it to arbitrary values of $K$.

\subsection{$S$-matrix and matrix Bethe equations}

Before addressing the $K=1$ case, we should first clarify that the twisted Hamiltonian (\ref{eq:twistedXXX}) does not commute with the regular permutation operator. Nevertheless, it commutes with the translation operator $U$ that we defined at the end of section 2, meaning that we can apply Bloch's theorem despite the twists. Although we are allowed to assume a plane-wave ansatz, we still have to include additional factors proportional to the twist parameters $q_i$. To illustrate this point, we show here the ansatz we will use for the case of two excitations with different flavour
\begin{align}\label{M2wavefunction}
	|\psi_{23} \rangle &=\sum_{1 \leq n_1 < n_2 \leq L} \left[ A_{1} e^{i(p_1  n_1 + p_2  n_2)} \frac{q_3^{n_1}}{q_2^{n_2}} + A_2 e^{i(p_1  n_2 + p_2  n_1)} \frac{q_3^{n_1}}{q_2^{n_2}} \right] S^{2,+}_{n_1} S^{3,+}_{n_2} |0\rangle \notag \\
	&+\sum_{1 \leq n_1 < n_2 \leq L} \left[ A_{3} e^{i(p_1  n_1 + p_2  n_2)} \frac{q_3^{n_2}}{q_2^{n_1}} + A_4 e^{i(p_1  n_2 + p_2  n_1)} \frac{q_3^{n_2}}{q_2^{n_1}} \right] S^{3,+}_{n_1} S^{2,+}_{n_2}|0\rangle \ ,
\end{align}
where $A_i$ are constants to be determined by requiring it to be an eigenvector of the Hamiltonian. The pseudo-vacuum $|0\rangle$ is constructed as the tensor product of $L$ states of type 1, i.e. $|0\rangle=\otimes_{n=1}^L |1\rangle$, and the operators $S_k^{n,+}$ transform a state of type $1$ in the $k$-th term of the tensor product into a state of type $n$.

The $S$-matrix for the excitations of the eclectic spin chain is the same as the $S$-matrix we find for the usual $SU(3)$ spin chain, with some additional twist factor in the diagonal entries
\begin{equation}
	S(p_2, p_1)=\begin{pmatrix}
		-\frac{1-2 e^{i p_2} + e^{i (p_1+p_2)}}{1-2 e^{i p_1} + e^{i (p_1+p_2)}} & 0 & 0 & 0 \\
		0 & \frac{1}{q_1 q_2 q_3} \frac{e^{i p_2}-e^{i p_1}}{1-2 e^{i p_1} + e^{i (p_1+p_2)}} & \frac{-(1-e^{i p_1})(1-e^{i p_2})}{1-2 e^{i p_1} + e^{i (p_1+p_2)}} & 0 \\
		0 & \frac{-(1-e^{i p_1})(1-e^{i p_2})}{1-2 e^{i p_1} + e^{i (p_1+p_2}} &  q_1 q_2 q_3 \frac{e^{i p_2}-e^{i p_1}}{1-2 e^{i p_1} + e^{i (p_1+p_2)}} & 0 \\
		0 & 0 & 0 & -\frac{1-2 e^{i p_2} + e^{i (p_1+p_2)}}{1-2 e^{i p_1} + e^{i (p_1+p_2)}}
	\end{pmatrix} \ .
\end{equation}
As the theory is integrable, we can write the matrix Bethe equations for any value of the number of excitations $M$ and $K$ as
\begin{equation}
	e^{i p_k L} q_3^{(3-f_k)L} q_2^{(2-f_k)L} |\psi \rangle =S(p_k , p_{k+1}) \dots S(p_k , p_M) S (p_k , p_1) \dots S (p_k , p_{k-1} ) |\psi \rangle \ ,
\end{equation}
where $f_k$ is the flavour of the $k$-th excitation. The unusual factor $q_3^{(3-f_k)L} q_2^{(2-f_k)L}$ appears due to the dressing of the momenta with the deformation parameters $q_i$. In particular, it gives us a dressing $q_2^{-L}$ if $f_k=3$ and a dressing $q_3^L$ if $f_k=2$. The presence of these additional factors prevents us from using the regular NCBA construction, but it is enough to slightly modify the prescription to make room for them.

Instead of considering the full matrix Bethe equations, it is simpler to solve first the auxiliary problem 
\begin{align}
	\lambda_k q_3^{(3-f_k)L} q_2^{(2-f_k)L} \left| \psi \right\rangle &=s(p_k , p_{k+1}) \dots s(p_k , p_M) s (p_k , p_1) \dots s (p_k , p_{k-1} ) \left| \psi \right\rangle \ ,  \label{auxeq} \\
	s(p_i,p_j) &=-\frac{1-2 e^{i p_j} + e^{i (p_i+p_j)}}{1-2 e^{i p_i} + e^{i (p_i+p_j)}} S(p_i,p_j) \ . \notag
\end{align}
Notice that the auxiliary problem has the same form as the matrix Bethe equation, the only difference being that we are using an ``undressed'' version of our original $S$-matrix. This auxiliary $S$-matrix is crafted to have its upper-left entry equal to $1$. This choice simplifies its action on a wavefunction that is composed mostly of excitations of type $2$, as we do not have to keep track of these factors when applying the $S$-matrix on our ansatz.

In addition, we will be changing from the momentum variable to a rapidity variable defined by the map $e^{ip_i}=\frac{u_i}{u_i+1}$, as it simplifies some intermediate steps. In these variables, the undressed $S$-matrix takes the form
\begin{equation}
	s(u_i,u_j)=\begin{pmatrix}
		1 & 0 & 0 & 0 \\
		0 & \frac{1}{q_1 q_2 q_3} \frac{u_j - u_i}{u_j-u_i+1} & \frac{1}{u_j-u_i+1} & 0 \\
		0 & \frac{1}{u_j-u_i+1} &  q_1 q_2 q_3 \frac{u_j - u_i}{u_j-u_i+1} & 0 \\
		0 & 0 & 0 & 1
	\end{pmatrix} \ . \label{reducedS}
\end{equation}

\subsection{Nested Coordinate Bethe Ansatz for $K=1$}

The case of $K=1$, i.e. the case of operators with exactly a single $\phi^3$, was analysed in detail in \cite{firstpart}. As a consequence, here we will only summarise the results and we refer the reader to said article for the details. We can solve the matrix Bethe equation for $K=1$ if we consider the following ansatz for the wavefunction with a general number of excitations
\begin{equation}
	\left| \psi \right\rangle =\sum_{\substack{1\leq n_i \leq L \\ n_i < n_{i+1}}} \sum_{k=1}^M \sum_{\sigma \in S_M} \psi_k (\sigma) e^{i \sum_j p_{\sigma (j)} n_j} \frac{q_3^{\sum_j n_j}}{(q_2 q_3)^{n_k}} S_{n_1}^{2+} \dots S_{n_{k-1}}^{2+} S_{n_{k}}^{3+} S_{n_{k+1}}^{2+} \dots S_{n_M}^{2+} |0\rangle\ . \label{wavefunctiongeneralM}
\end{equation}
Here we will only be interested in the coefficients $\psi_k (Id.)$ because the remaining $\psi_k (\sigma)$ can be related to the former by translations. Thus, to alleviate our notation, in the following we will write the coefficient $\psi_k (Id.)$ as just $\psi_k$ unless it is necessary to indicate it.

Substituting this ansatz into the auxiliary equation (\ref{auxeq}), we find that the coefficients $\psi_k$ have to satisfy the equations
\begin{gather}
	q_3^L \lambda_k \psi_{k-l-1}=\frac{1}{u_{k-l-1}-u_{k}+1} \psi^{(l-1)}_k + q_1 q_2 q_3 \frac{u_{k-l-1}-u_{k}}{u_{k-l-1}-u_{k}+1} \psi_{k-l-1} \ , \label{Suth1} \\
	\psi^{(l+1)}_k=\frac{1}{q_1 q_2 q_3} \frac{u_{k-l-1}-u_{k}}{u_{k-l-1}-u_{k}+1} \psi^{(l)}_k +\frac{1}{u_{k-l-1}-u_{k}+1} \psi_{k-l-1} \ , \label{Suth2}
\end{gather}
where $\psi^{(l)}_k$ are some auxiliary coefficients with $\psi^{(-1)}_k=\psi_k$. These auxiliary coefficients have a well-defined physical interpretation. From the perspective of computing the Bethe equations by picking a particular excitation (which in this case we pick to be the only excitation of type $3$), moving it around the spin chain and scattering it with every other excitation until we put it back to its original place, the coefficients $\psi^{(l)}_k$ are the intermediate state of this excitation after interacting with $l+1$ of the other excitations.

The above equations can be recast into the following recurrence relations
\begin{gather}
\frac{\psi_{k-l-1}}{\psi_{k-l}} = \frac{1}{q_1 q_2 q_3} \, \frac{u_{k-l} - \bar{x}}{u_{k-l-1} - \bar{x}+1} \ , \quad  \psi^{(l-1)}_k =q_1 q_2 q_3 \frac{u_{k-l-1} - \bar{x} +1}{u_k - \bar{x}} \psi_{k-l-1} \ , \label{Suth} \\
\text{ with } \quad \bar{x} -u_k = \frac{q_1 q_2 q_3}{q_1 q_2 q_3- \lambda_k^{(1)} q_3^{L}}\ . \notag
\end{gather}
Here we have added the superindex $(1)$ to the eigenvalue of the auxiliary equation $\lambda_k$ to indicate that we are in the case of $K=1$. As the left-hand-side of the first equation depends only on the difference $k-l$ while the right-hand-side depends explicitly on $k$ (through $\bar{x}$), this solution to the auxiliary equations is consistent only if $\bar{x}$ is a constant. This constant plays the rôle of auxiliary rapidity. Identifying the excitation $M+1$ with the excitation $1$ gives us the following constraint for the single auxiliary rapidity of this problem
\begin{equation}
	\frac{\lambda_k^{(1)}}{q_2^L} \prod_{\substack{j=1 \\ j\neq k}}^M q_3^L \lambda_j^{(1)} =1 \Longrightarrow  \frac{({q_2}{q_3})^L}{({q_1}{q_2}{q_3})^{M}}
\prod_{j=1}^M\frac{\bar{x}-u_j}{\bar{x}-u_j-1}=1 \ . \label{AuxBetheEq}
\end{equation}
This equation for $\bar{x}$ is the auxiliary Bethe equation of our model. An important point we want to emphasise here is that the recurrence equation of the coefficients $\psi_k$ is not as general as it might look. In particular, we should stress that there is a coefficient that plays a special rôle. The coefficient $\psi_k$, with the same subindex as the momentum of the Bethe equation $p_k$ we are looking at, behaves differently than the other coefficients due to the twist factors. This means that the recurrence equation is only valid for $l$ between $0$ and $M-2$, while the relation for $M-1$ has extra twist factors, which account for the additional $({q_2}{q_3})^L$ factor in the auxiliary Bethe equations from the naïve expectation.

In addition, if we substitute the expression for the auxiliary coefficients (\ref{Suth}) into the equation (\ref{Suth2}), we can see that these coefficients satisfy the relation
\begin{equation}
	q_3^L \lambda_{j}^{(1)} \psi_{k-l-1}=q_1 q_2 q_3 \frac{u_{k-l-1}-u_{k}}{u_{k-l-1}-u_{k}+1} \psi_{k-l-1} +\frac{1}{u_{k-l-1}-u_{k}+1} \frac{\psi_k}{(q_3^L \lambda_{k-1}^{(1)}) \dots (q_3^L \lambda_{k-l}^{(1)})} \ . \label{Recurrencepsi}
\end{equation}
This equation will be extremely useful for the $K=2$ case.

Finally, once we have solved the matrix Bethe equations for the undressed $S$-matrix, solving the matrix Bethe equations for the original $S$-matrix is immediate, as they reduce to the algebraic equation
\begin{equation}
	e^{i p_k L}= \lambda_k^{(1)} \prod_{j \neq k} \left( - \frac{1-2 e^{i p_k} + e^{i (p_k+p_j)}}{1-2 e^{i p_j} + e^{i (p_k+p_j)}} \right)
\end{equation}
where $\lambda^{(1)}_k$ are the eigenvalues we have computed above in terms of $\bar{x}$. Substituting them and replacing the momenta by rapidities, the Bethe equations takes the form
\begin{equation}
	\frac{q_3^L}{q_1 q_2 q_3} \frac{\bar{x} -u_k}{\bar{x} -u_k-1} \prod_{j \neq k} \frac{u_k - u_j +1}{u_k - u_j -1} = \left( \frac{u_k+1}{u_k} \right)^L \ . \label{BetheEq}
\end{equation}
The fact that both the Bethe equations and the auxiliary Bethe equation we obtained are exactly the same as the ones appearing in \cite{StaudacherAhn} when we identify our $\bar{x}$ with their $v+1$ works as a consistency check.

\subsection{Nested Coordinate Bethe Ansatz for $K=2$}

For the case involving two excitations of type $3$, i.e. the $K=2$ case, we have to consider the following wavefunction
\begin{align}
	\left| \psi \right\rangle &=\sum_{\substack{1\leq n_i \leq L \\ n_i < n_{i+1}}} \,  \sum_{\substack{j,k=1 \\ j<k}}^M \sum_{\sigma \in S_M} \psi_{jk} (\sigma) e^{i \sum_l p_{\sigma (l)} n_l} \frac{q_3^{\sum_l n_l}}{(q_2 q_3)^{n_j + n_k}} S_{n_1}^{2+} \dots  S_{n_{j-1}}^{2+} S_{n_{j}}^{3+} S_{n_{j+1}}^{2+} \dots \notag \\
	&\dots S_{n_{k-1}}^{2+} S_{n_{k}}^{3+} S_{n_{k+1}}^{2+} \dots S_{n_M}^{2+} |0\rangle \ , \label{wavefunctiongeneralM2}
\end{align}
together with the ansatz that the coefficients $\psi_{jk}$ can be expressed in terms of the $K=1$ coefficients (\ref{Suth}) as
\begin{equation}
	\psi_{jk} (\sigma , \bar{x} , \bar{x}')= B \psi_j (\sigma , \bar{x}) \psi_k (\sigma , \bar{x}') + \tilde{B} \psi_k (\sigma , \bar{x}) \psi_j (\sigma , \bar{x}') \ ,
\end{equation}
where we have explicitly written the dependence on the two auxiliary rapidities. To alleviate notation, we will indicate that a quantity depends on the auxiliary rapidity $\bar{x}'$ by adding a prime to the quantity, while quantities with no prime should be understood as functions of $\bar{x}$.

Thanks to this ansatz, computing the action of the auxiliary matrix equation (\ref{auxeq}) on the wavefunction simplifies enormously. For example, let us consider the action of the first undressed $S$-matrix in the auxiliary problem, $ s (p_k , p_{k-1} )$, when the two excitations of type $3$ are well-separated. The equations we find in this situation are exactly the same ones we had for $K=1$
\begin{equation}
	q_3^L \lambda_k \psi_{j,k-1}=\frac{1}{u_{k-1}-u_{k}+1} \psi_{j,k} + q_1 q_2 q_3 \frac{u_{k-1}-u_{k}}{u_{k-1}-u_{k}+1} \psi_{j,k-1} \ ,
\end{equation}
so they are satisfied by our ansatz. On the other hand, if both excitations $k-1$ and $k$ are of type $3$, the coefficient is unaltered, as the lower-right coefficient of the reduced $S$-matrix (\ref{reducedS}) is equal to $1$.

Let us now go one step forward, and consider the action of $ s (p_k , p_{k-2} ) s (p_k , p_{k-1} )$ on the wavefunction and study the coefficient we find for the case where the excitations $k-2$ and $k-1$ are both of type $3$. According to the above discussion, the action of $s (p_k , p_{k-1} )$ transform the coefficient $\psi_{k-2,k-1}$ into $B \psi_{k-2} \psi'_{k-1} (q_3^L \lambda'_{k}\null^{(1)}) + \tilde{B} (\bar{x} \leftrightarrow \bar{x}')$, while it leaves $\psi_{k-1,k}$ unaltered. It may not look like that at first sight, but this poses a problem. If we consider the recurrence relation the $\psi_k$ functions fulfil (\ref{Recurrencepsi}), we can see that the coefficients we find after acting with $s (p_k , p_{k-1} )$ do not have the correct form for $\psi_{jk}$ to be a solution of the auxiliary equation, in the sense that applying now $s (p_k , p_{k-2} )$ gives us
\begin{equation}
	\frac{1}{u_{k-1}-u_{k}+1} \tilde{B} \psi'_{k-1} \psi_{k}+ q_1 q_2 q_3 \frac{u_{k-1}-u_{k}}{u_{k-1}-u_{k}+1} B \psi_{k-2} \psi'_{k-1} (q_3^L \lambda'_{k}\null^{(1)}) + (\bar{x} \leftrightarrow \bar{x}') \ ,
\end{equation}
so we cannot recast the coefficient associated to the case where both $k-2$ and $k-1$ are of type $3$ into something proportional to $\psi_{k-2,k-1}$. If we want to get the correct eigenvalue equation, we have to find a way to substitute $\psi_{k-1,k}$ by $B \psi'_{k-1} \psi_k (q_3^L \lambda'_{k}\null^{(1)}) (q_3^L \lambda_{k-1}^{(1)})^{-1} + \tilde{B} \psi_{k-1} \psi'_k (q_3^L \lambda_{k}^{(1)})(q_3^L \lambda'_{k-1}\null^{(1)})^{-1}$.

The only way to solve this mismatch between our expectations and the actual calculations is to choose the coefficients $B$ and $\tilde{B}$ in such a way that
\begin{displaymath}
	B \psi_{k-1} \psi'_k + \tilde{B} \psi'_{k-1} \psi_k= B \psi'_{k-1} \psi_k (q_3^L \lambda'_{k}\null^{(1)}) (q_3^L \lambda_{k-1}^{(1)})^{-1} + \tilde{B} \psi_{k-1} \psi'_k (q_3^L \lambda_{k}^{(1)})(q_3^L \lambda'_{k-1}\null^{(1)})^{-1} \ .
\end{displaymath}
Once we fix them as such, we can use equation (\ref{Recurrencepsi}) with no problem. As no other undressed $S$-matrix acts non-trivially again on neither the site $k-1$ nor the site $k-2$, this implies that $\psi_{k-2,k-1}$ is the correct wavefunction and it is associated to the eigenvalue
\begin{equation}
	q_3^L \lambda_{k}^{(2)} = q_3^{2L} \lambda_{k}^{(1)} \lambda'_{k}\null^{(1)} \ .
\end{equation}

Although we have solved the problem for the action of two undressed $S$-matrices, the exact same problem will appear in all the subsequent cases. In particular, after acting with $l$ of these undressed $S$-matrices, we find that the coefficients $\psi_{jk}$ give us the correct eigenvectors only if $B$ and $\tilde{B}$ satisfy the following relation
\begin{multline}
	B \psi_{k-l} \psi'_k (q_3^L \lambda'_{k-1}\null^{(1)})^{-1} \dots (q_3^L \lambda'_{k-l+1}\null^{(1)})^{-1} + \tilde{B} (\bar{x}\leftrightarrow\bar{x}')=\\
	= B \psi'_{k-1} \psi_k (q_3^L \lambda'_{k}\null^{(1)})  (q_3^L \lambda_{k-1}^{(1)})^{-1} \dots (q_3^L \lambda_{k-l}^{(1)})^{-1} + \tilde{B} (\bar{x}\leftrightarrow\bar{x}') \ .
\end{multline}
It may look like our situation is dire, as this equation has to be satisfied for every value of $l$ and we have only two free coefficients. Luckily, after substituting the explicit expression for the eigenvalues $\lambda^{(1)}$, we get
\begin{equation}
	\frac{\tilde{B}}{B}=\frac{\bar{x}'-\bar{x} -1}{\bar{x}'-\bar{x}+1} \ ,
\end{equation}
which is independent of $l$, meaning that all the equations are consistent and can be solved simultaneously.

As in the $K=1$, we get the auxiliary Bethe equation by identifying the excitation $M+1$ with the excitation $1$. Let us consider the case of $k=M$ for simplicity, where the identification implies solving the equation
\begin{equation}
	B \frac{\psi_{M} \psi'_{M-j} (q_3^L \lambda'_{M}\null^{(1)}) }{(q_3^L \lambda_{M-1}^{(1)}) \dots (q_3^L \lambda_{1}^{(1)})} + \tilde{B} (\bar{x}\leftrightarrow\bar{x}') = q_2^{-L} \lambda_{M}^{(2)} \left[ B \psi'_{M} \psi_{M-j} + \tilde{B} (\bar{x}\leftrightarrow\bar{x}')  \right] \ .
\end{equation}
Equating the $B$ term of the left-hand-side with the $\tilde{B}$ term of the right-hand-side, we find the following constraint
\begin{equation}
	\frac{B}{\tilde{B}} = \frac{\prod_{j=1}^M (q_3^L \lambda_{j}^{(1)})}{q_2^L q_3^L} =  \frac{q_2^L q_3^L}{\prod_j (q_3^L \lambda'_{j}\null^{(1)})} \ .
\end{equation}
Substituting the expression we found for the ratio of the coefficients and the eigenvalues $\lambda_{j}^{(1)}$, we find the following auxiliary Bethe equation
\begin{equation}
	\frac{q_1^M q_2^M q_3^M}{q_2^L q_3^L} \prod_{j=1}^M \frac{u_j - \bar{x} +1}{u_j - \bar{x}}= \frac{\bar{x} - \bar{x}'-1}{\bar{x} - \bar{x}'+1} \ . \label{auxBetheK2}
\end{equation}

The only thing left to do is to use this solution of the auxiliary equation to solve the matrix Bethe equation. If we substitute the eigenvalue $\lambda_k^{(2)}$ into the relation
\begin{displaymath}
	e^{i p_k L}= \lambda_k^{(2)} \prod_{j \neq k} \left( - \frac{1-2 e^{i p_k} + e^{i (p_k+p_j)}}{1-2 e^{i p_j} + e^{i (p_k+p_j)}} \right) \ ,
\end{displaymath}
and change the momenta into rapidities, it becomes
\begin{equation} \frac{q_3^L}{(q_1 q_2 q_3)^2} \frac{\bar{x} -u_k}{\bar{x} -u_k-1} \frac{\bar{x}' -u_k}{\bar{x}' -u_k-1} \prod_{j \neq k} \frac{u_k - u_j +1}{u_k - u_j -1} = \left( \frac{u_k+1}{u_k} \right)^L \ . \label{BetheEqK2}
\end{equation}
The fact that, again, both the Bethe equations and the auxiliary Bethe equations we obtained are exactly the same as the ones appearing in \cite{StaudacherAhn} when we identify our $\bar{x}$ with their $v+1$ works as a consistency check.

\subsection{Nested Coordinate Bethe Ansatz for general $K$}

The method we presented for constructing the $K=2$ wavefunction using the $K=1$ wavefunctions as building blocks can be upgraded to work for general values of $K$.\footnote{As Bill Sutherland masterfully summarises it before his explanation \cite{Sutherland}, ``Is there any portion of this scheme which doesn't immediately generalise to more than two overturned spins? No.''}

As we did in the case of $K=2$, we assume a wavefunction of the form
\begin{align}
	\left| \psi \right\rangle &= \sum_{\substack{1\leq n_i \leq L \\ n_i < n_{i+1}}}  \, \sum_{\substack{1\leq k_i \leq M \\ k_i < k_{i+1}}}  \sum_{\sigma \in S_M} \psi_{k_1 , k_2 , \dots , k_K} (\sigma) e^{i \sum_l p_{\sigma (l)} n_l} \frac{q_3^{\sum_l n_l}}{(q_2 q_3)^{\sum_i n_{k_i}}} S_{n_1}^{2+} \dots S_{n_{k_1}-1}^{2+}  S_{n_{k_1}}^{3+} S_{n_{k_1}+1}^{2+} \dots \\
	&\dots S_{n_{k_K}-1}^{2+} S_{n_{k_K}}^{3+} S_{n_{k_K}+1}^{2+} \dots S_{n_M}^{2+} |0\rangle\ , \label{wavefunctiongeneralMK}
\end{align}
where the coefficient $\psi_{k_1 , k_2 , \dots , k_K}$ can be written as a linear combination of products of solutions of the $K=1$ case
\begin{equation}
	\psi_{k_1 , k_2 , \dots , k_K} (\sigma)= \sum_{\rho \in S_K} B(\rho) \prod_{j=1}^K \psi_{k_j} (\sigma , \bar{x}_{\rho(j)}) \ .
\end{equation}
Here again we have written the dependence of $\psi_k$ on the auxiliary rapidity explicitly. By considering the case where two excitations are close together and the other ones are well-separated, it is evident that the coefficients will satisfy similar equations to the ones of the $K=2$ case. In particular
\begin{equation}
	\frac{B(\tau_{j,j+1} \rho)}{B(\rho)}= \frac{\bar{x}_{j+1}-\bar{x}_j -1}{\bar{x}_{j+1}-\bar{x}_j+1}
\end{equation}
where $\tau_{j,j+1}$ is the permutation of $k_j$ and $k_{j+1}$. In addition, the eigenvalues of the auxiliary equation become nothing else but the products of the eigenvalues for $K=1$ for each auxiliary rapidity
\begin{displaymath}
	q_3^L \lambda_k^{(K)}=q_3^{KL} \prod_{j=1}^{K} \lambda_{k} \null^{(1)} (\bar{x}_j) \ ,
\end{displaymath}
while the identification of the excitation $M+1$ with the excitation $1$ takes the form
\begin{equation}
	\frac{B(\prod_{m=n+1}^{n-1} \tau_{m,n})}{B(Id.)}= \frac{\prod_{j=1}^M [q_3^L \lambda_{j}^{(1)} (x_n)]}{q_2^L q_3^L} \ ,
\end{equation}
which is nothing but the auxiliary Bethe equation
\begin{equation}
	\prod_{m \neq n} \frac{\bar{x}_m - \bar{x}_n -1}{\bar{x}_m - \bar{x}_n +1}=\frac{(q_2 q_3)^L}{(q_1 q_2 q_3)^M} \prod_j \frac{\bar{x}_n - u_j}{\bar{x}_n - u_j - 1} \ . \label{auxBetheKK}
\end{equation}

Finally, substituting these results into the matrix Bethe equation, we find the following Bethe equation
\begin{equation}
\frac{q_3^L}{(q_1 q_2 q_3)^K} \prod_{l=1}^K \frac{\bar{x}_l -u_k}{\bar{x}_l -u_k-1} \prod_{j \neq k} \frac{u_k - u_j +1}{u_k - u_j -1} = \left( \frac{u_k+1}{u_k} \right)^L \ . \label{BetheEqKK}
\end{equation}
As in the previous cases, these equations match the ones appearing in \cite{StaudacherAhn} when we identify our $\bar{x}$ with their $v+1$ works as a consistency check.

\section{Coalescence of the Bethe states and the structure of generalised eigenvectors}

In this section, we will construct the generalised eigenvectors of the eclectic spin chain by means of the method described in section 3. This means that we will compute the limit of large twist of linear combinations of the wavefunctions we constructed in section 4. We will show that, as it happens in the $K=1$ case studied in our previous article \cite{firstpart}, only a small set of coefficients contribute in this limit. This result simplifies the calculation we have to perform, and we are able to identify how many states we find at a given step of the process of extracting the generalised eigenvectors by finding the number of integer solutions of a system of linear equations.

If we want to find how each of the terms appearing in the wavefunction (\ref{wavefunctiongeneralMK}) behaves in the strongly twisted limit, we need to find how the momenta $p_j$ and the coefficients $\psi_{k_1 , \dots , k_K}$ behave in such a limit.

For the momenta, we have to study first the behaviour of the Bethe equations (\ref{BetheEqKK}) and the auxiliary Bethe equation (\ref{auxBetheKK}) in the limit of large twist. This computation was carried out previously in \cite{StaudacherAhn}, so we can just borrow their results, which holds for $L>3(M-K)$
\begin{equation}
	u_j \approx \epsilon^\alpha u^-_j \ , \qquad u_{M-K+l} \approx -1+\epsilon^\beta u^+_l \ , \qquad \bar{x}_l \approx u_{M-K+l} + \epsilon^\gamma \hat{v} \ ,
\end{equation}
where $\alpha = \frac{L-M-K}{L-M+K}$,  $\beta = \frac{L-3M+3K}{L-M+K}$ and $\gamma = 2 L-3M-(K-1) \beta$. Notice that the requirement $L>3(M-K)$ prevents the coefficient $\beta$ from being negative. In addition, requiring that the number of excitations type $2$ is larger than the number of excitations of type $3$ implies that $\alpha > \beta$. After substituting this asymptotic behaviour, the Bethe equations take the form
\begin{align}
	&(u^-_j)^L =(-1)^{M-1} \frac{\xi^K}{\xi_3^L} \prod_{l=1}^K u^+_l \ ,  \label{Bethelimit} \\
	&(-u^+_l)^{L-M+K} = (-1)^K \frac{\xi_3^L}{2^{M-K} \xi^K} \hat{v}_l \prod_{k\neq l} (u^+_l - u^+_k) \ , \\
	&-1=\frac{\xi^{L-M}}{2^{M-K} \xi_1^L} \hat{v}_l \prod_{k\neq l} (u^+_l - u^+_k) \ ,
\end{align}
where $\xi=\xi_1 \xi_2 \xi_3$. At the level of momenta, the behaviour of the rapidities translates into
\begin{equation}
	e^{i p_j} = \frac{u^-_j \epsilon^\alpha}{u^-_j \epsilon^\alpha + 1} \approx u^-_j \epsilon^\alpha \ , \qquad e^{i p_{M-K+l}} = \frac{-1+\epsilon^\beta u_{l}^+}{\epsilon^\beta u_{l}^+} \approx -(u_{l}^+)^{-1} \epsilon^{-\beta} \ .
\end{equation}

Let us consider now the coefficients $\psi_{i}$. Substituting the behaviour of the rapidities into the relation (\ref{Suth}), we find that
\begin{equation}
	\frac{\psi_{k-l-1} (Id.)}{\psi_{k-l} (Id.)} = \frac{1}{q_1 q_2 q_3} \, \frac{u_{k-l} - \bar{x}}{u_{k-l-1} - \bar{x}+1}= \begin{cases} \frac{1}{2} \epsilon^{3} 
	& \text{ if } k-l \leq M-K \\
	\frac{ u^+_{k-l} - u^+_j}{2} \epsilon^{\beta+3} & \text{ if } k-l=M-K-1 \\
	(u^+_{k-l} - u^+_j) \epsilon^{\beta+3} & \text{ if } k-l > M-K-1  \end{cases} \ .
\end{equation}
As the leading order does not depend on the auxiliary rapidities in any of the three cases, the recurrence relation extend immediately to the coefficients with several indices $\psi_{k_1 , k_2 , \dots , k_K}$. In addition, moving away from a coefficient where all the excitations of type $3$ are placed together and to the right increases the power of $\epsilon$. This means that the leading order coefficients are always of the form $\psi_{{M-K+1},{M-K+2},\dots , M}  (Id.)$ or one obtained from it after using the translation operator $U$ enough times.

The final piece we need to understand the strongly twisted limit of the wavefunctions is a subtlety due to the form of the solutions of the Bethe equation. By direct inspection of (\ref{Bethelimit}), we can see that $M-K$ of the Bethe roots $u_j$ become the same in the strongly twisted limit. This poses a problem, as two coefficients $\psi (\sigma)$ and $\psi (\sigma ')$ that are related by a transposition that involves two particles with the same flavour will be equal up to a sign, making the wavefunction vanish at leading order. Luckily, the next-to-leading order contribution to the Bethe roots is different for each rapidity. If we substitute the approximation $u_j \approx u_j^- \epsilon^\alpha + \tilde{u}^-_j \epsilon^{2\alpha} + \dots$ into the Bethe equations, we find that
\begin{equation}
	\frac{(u^-_j)^2 - \tilde{u}^-_j}{(u^-_j)^{L+1}} = 2 \frac{\xi_3^L}{\xi^K} \frac{\sum_{k\neq j} (\tilde{u}^-_j - \tilde{u}^-_k)}{\prod_{l=1}^K u^+_l} \ ,
\end{equation}
which implies that the $\tilde{u}^-_j$ are all different. Nevertheless, this subtlety does not affect the previous discussion, and the contribution associated to $\psi_{{M-K+1},{M-K+2},\dots , M}  (Id.)$ is still the dominant contribution to the wavefunction.

Putting everything we discussed together, the wavefunction (\ref{wavefunctiongeneralMK}) close to the limit of strong twist can be approximated by
\begin{align}
	\left| \psi \right\rangle &\approx \sum_{\substack{1\leq n_i \leq L \\ n_i < n_{i+1}}} \psi_{{M-K+1},{M-K+2},\dots , M} (Id.) e^{i \sum_l p_{\sigma (l)} n_l} \frac{q_3^{\sum_{l=1}^{M-K} n_l}}{q_2^{\sum_{l=M-K+1}^M n_{l}}} \cdot \notag \\
	&\cdot S_{n_1}^{2+} S_{n_2}^{2+} \dots  S_{n_{M-K}}^{2+} S_{n_{M-K+1}}^{3+} \dots S_{n_M}^{3+} |0\rangle + \text{translations} \ .
\end{align}

As the coefficient $\psi_{{M-K+1},{M-K+2},\dots , M}  (Id.)$ does not depend on the position of the excitations, $n_i$, the only part of the wavefunction that plays a rôle into selecting which vector survives in the strongly twisted limit is the plane wave factor. Substituting the expression for the momenta, the plane wave factor at strong twist behaves as
\begin{equation}
	\left( e^{i \sum p_i n_i} \frac{q_3^{\sum_{i=1}^{M-K} n_i}}{q_2^{\sum_{j=1}^K n_{M-K+j}} } \right)^K \sim \epsilon^{K(\alpha-1) \sum_{i=1}^{M-K} n_i - K(\beta -1) \sum_{j=1}^K n_{M-K+j}}= \epsilon^{ (\alpha-1) \sum_{i=1}^{M-K} \sum_{j=1}^K ( n_i - n_{M-K+j} ) } \ ,
\end{equation}
where we made use of the relation $(M-K) (\alpha-1)=K(\beta -1)$. We should remark that this expression depends only on the relative position of the excitations of type $2$ and $3$. This was not unexpected, as the states we constructed are eigenvectors of the translation operator $U$.

Thus, if we compute the strongly twisted limit of the wavefunction, the only terms that appear are those that make the sum $\sum_{i=1}^{M-K} \sum_{j=1}^K ( n_{M-K+j} - n_i )$ minimal. Notice that we have changed the sign of the addends because both $\alpha - 1 $ and $n_i - n_{M-K+j}$ are negative. Accomplishing this requirement is incredibly easy, as we just have to put all the excitations as close together as they can be, giving us the locked state we discussed in section~\ref{themodel}. This means placing the excitations at $n_i=L-M+i$, up to translations.

As we described in \cite{firstpart} and reviewed in section \ref{themethod}, the limit of a linear combination of eigenvectors contains information about the generalised eigenvector of rank two in the non-diagonalizable limit. In this setting, this implies that the vectors associated to the generalised eigenvector of rank two are those associated to the terms that makes the sum $\sum_{i=1}^{M-K} \sum_{j=1}^K ( n_i - n_{M-K+j} )$ as close as possible to minimal, but without reaching the minimal value. This can be accomplished in this case by taking the eigenvector we have just computed and moving the excitation $n_1$ one site to the left while keeping the other ones fixed.\footnote{As excitations are tightly packed, we can only move either the excitation of type $2$ at $n_1=L-M+1$ or the excitation of type $3$ at $n_M=L$ (up to translations). While moving the excitation of type $2$ increases the sum by $K$, moving the excitation of type $3$ increases the sum by $M-K>K$. Thus, the former is the next-to-minimal solution.}

The process is the same for generalised eigenvectors of higher rank. Thus, we can compute the number of independent generalised eigenvectors that we would have found at every step of the process by calculating the number of solutions to the Diophantine equation (as we are only interested in integer solutions of this equation)
\begin{equation}
	\sum_{i=1}^{M-K} \sum_{j=1}^K ( n_{M-K+j} - n_i ) = n \quad \text{ with } \quad 1 \leq n_1 < n_2 <\dots < n_M \leq L \ , \label{Diophantine}
\end{equation}
and multiplying it by the length of the spin chain $L$. From this result, the length of the Jordan chains can be obtained by computing how many values of $n$ have a given number of solutions or more.

This additional factor of $L$ comes from the fact the wavefunction still depends on the total momentum, given by $e^{i \sum_{j=1}^M p_j} \approx \frac{\prod_{j=1}^{M-K} u_j^-}{\prod_{j=1}^K u_j^+} \epsilon^{\zeta}$, where the value of $\zeta$ is not important for our discussion. By examining the Bethe equation (\ref{Bethelimit}), we can check that the total momentum can take $L$ different values. We can clearly see this dependence on the total momentum in the locked state we wrote (\ref{lockedstate}).

However, we cannot start computing the number of solutions of this equation yet, as this recipe is not enough to cover the full Hilbert space except for $K=1$. This is immediate to see, as translations of $\psi_{{M-K+1},{M-K+2},\dots , M} (Id.)$ cannot create a non-vanishing contribution to the coefficient $\psi_{{M-K},{M-K+2},\dots , M}  (Id.)$ unless $K=1$. This means that we can never put an excitation of type $2$ between two excitations of type $3$. Thus, we have to consider the contribution from some additional coefficients. The same rules we followed before to argue that $\psi_{{M-K+1},{M-K+2},\dots , M} (Id.)$ and translations of it are the leading order coefficients while the remaining ones are negligible in comparison can be applied here. Thus, we have to choose all the coefficients with the excitations as close together as possible, with the excitations of type $3$ as close to the right as possible, and inequivalent between them under translations. This way, we can be sure that all coefficients of the wavefunction are non-trivial, and that the different coefficients we have chosen do not interfere with each other. The practical implications of taking them into account is that we have to consider all the orderings of the excitations of type $2$ and $3$ that are not equivalent under translations. For example, for $M=5$ and $K=2$ we have to consider the coefficients $\psi_{45} (Id.)$ and $\psi_{35} (Id.)$ (and those obtained from them by translations).

After this observation, we would have to repeat all the same arguments for the additional coefficients we have to take into account. Luckily, the same reasoning as before can be applied mutatis mutandis to these additional contributions, so we can compute the number of independent generalised eigenvectors we find at a given step by solving the similarly looking Diophantine equation
\begin{equation}
	\sum_{j\in \text{type }2} \sum_{k\in \text{type }3}  (n_k - n_j) = n \ . \label{Diophgeneral}
\end{equation}
Although some of the terms in this sum are negative, as we can have excitations of type $2$ to the right of excitations of type $3$, it is easy to see that the total sum will always give us a positive number.

\section{Solving the Diophantine equation}

In this section, we will analyse the number of integer solutions of the equation (\ref{Diophgeneral}) for the different orderings of excitations we have to consider. We will begin by reviewing the $K=1$ case, as we will use the structures we find there as the building blocks for the other cases.

\subsection{The Diophantine equation in the $K=1$ case}

For $K=1$, the only Diphantine equation we have to consider is equation (\ref{Diophantine})
\begin{displaymath}
	\sum_{j=1}^{M-1} (n_M - n_j) = n \quad \text{ with } 1 \leq n_1 < n_2 <\dots < n_M \leq L \ .
\end{displaymath}
As this equation only depends on the difference of the positions of the excitations, it simplifies heavily if we change our variables from the position of the excitations $n_i$ to the relative position of two adjacent excitations $x_i=n_{i+1} - n_i+1$. In these variables, the previous equation becomes
\begin{displaymath}
	\sum_{j=1}^{M-1} j x_j = \Delta \quad \text{ with } x_i \geq 0 \text{ and } \sum_{j=1}^{M-1} x_j \leq L-M\ ,
\end{displaymath}
where $\Delta=n-\frac{M (M+1)}{2}$ is a more convenient variable. In addition, we can introduce the dummy variable $x_0$ in order to transform the last inequality into the linear equation $\sum_{j=0}^{M-1} x_j = L-M$. Therefore, we have to find the number of integer solutions of the following standard system of linear equations
\begin{equation}
	\left\{ \begin{matrix}
	\sum_{j=0}^{M-1} j x_j = \Delta \\
	\sum_{j=0}^{M-1} x_j = L-M \\
	x_j \geq 0 \ \forall j
\end{matrix}	 \right. \ . \label{simplifieddiophantine1}
\end{equation}

The key to finding the number of solutions to this equation is to notice that a set of numbers $(x_0 +1 , x_1 , \dots x_{M-1})$ is a solution for a given length $L+1$, number of excitations $M$ and sum $\Delta$ if $(x_0 , x_1 , \dots x_{M-1})$ is a solution for a length $L$, number of excitations $M$ and sum $\Delta$. The only solutions we cannot get through this recursive procedure are those with vanishing $x_0$.

If we substitute $x_0=0$ in the system (\ref{simplifieddiophantine1}), and subtract the second equation from the first one once, we find
\begin{equation}
	\left\{ \begin{matrix}
	\sum_{j=1}^{M-1} (j-1) x_j = \Delta + M -L \\
	\sum_{j=1}^{M-1} x_j = L-M \\
	x_j \geq 0 \ \forall j
\end{matrix}	 \right. \ ,
\end{equation}
which is exactly the Diophantine equation we would find for length $L-1$ and $M-1$ excitations.

This recursive method for constructing solutions is more appealing if we write it in terms of the generating function of the number of solutions. Consider
\begin{equation}
	F(L,M,K,q)=\sum_{\Delta=0}^\infty q^\Delta \# \{\text{Solutions of eq.~\ref{simplifieddiophantine1} for given values of } L,M,K,\Delta\} \ .
\end{equation}
In terms of this generating function, the recurrence relation we have found can be written as
\begin{equation}
	F(L,M,1,q)=F(L-1,M,1,q)+q^{L-M} F(L-1,M-1,1,q) \ ,
\end{equation}
which has to be supplemented with the initial condition $F(L,2,1,q)=\sum_{j=0}^{L-2} q^j$. This condition comes from noticing that for $M=2$ we have two equations and two unknowns, meaning that we have exactly one solution for every admissible value of $\Delta$ (that is, every integer from $0$ up to $L-2$).

The solution to this recurrence relation is a known family of functions called \emph{Gaussian polynomials} or \emph{$q$-deformed binomial coefficients}. In particular, our generating function corresponds to
\begin{equation}
	F(L,M,1,q)=\binom{L-1}{M-1}_q = \prod_{k=1}^{M-1} \frac{1-q^{L-k}}{1-q^k} \ .
\end{equation}

As we commented in the previous section, the number of generalised eigenvectors is equal to the number of allowed values of the total momentum times the number of solutions of the Diophantine equation. As we have a total of $L$ possible values of the total momentum, the actual number of solutions is given by $L\binom{L-1}{M-1}_q$. This result was first derived using combinatorial arguments in \cite{Ahn:2021emp} and shortly after using the method detailed in section 3 in \cite{firstpart}.

\subsection{The Diophantine equation in the $M=4$ and $K=2$ case}

Let us move now to the simplest setting with $K>1$, which consist of two excitations of type $2$ and two excitations of type $3$.

As we commented in the previous section, we need to consider all the ways of arranging the excitations that are not equivalent under translations of all the sites of the chain. In this case we have two of them: either $\psi_{34} (Id.)$, corresponding to the arrangement $2233$, or $\psi_{24} (Id.)$, corresponding to the arrangement $2323$. The Diophantine equations for these arrangements are, respectively,
\begin{displaymath}
	(n_4 - n_1) + (n_4 - n_2) + (n_3 - n_1) + (n_3 - n_2) = n \text{ and } (n_4 - n_1) + (n_4 - n_3) + (n_2 - n_1) + (n_2 - n_3) = n \ .
\end{displaymath}
Notice that, despite $n_2 - n_3$ being negative in the second equation, the whole sum is always positive because $n_4 -n_1 > n_3 - n_2$. This fact is even clearer when we write the equations in terms of the relative positions $x_i$
\begin{equation}
	\left\{ \begin{matrix}
	2x_1 + 4 x_2 + 2 x_3 = n-8= 2 \Delta\\
	\sum_{j=0}^{3} x_j = L-4 \\
	x_j \geq 0 \ \forall j
\end{matrix}	 \right.  \text{ and } \left\{ \begin{matrix}
	2x_1 + 2 x_3 = n-4= 2 \Delta\\
	\sum_{j=0}^{3} x_j = L-4 \\
	x_j \geq 0 \ \forall j
\end{matrix}	 \right.  \ .
\end{equation}

Although these two systems of equations resemble the one we had for $K=1$, they are not of the same form. The first system of equations would be the same as the one for $K=1$ if we were able to consider the variables $x_1$ and $x_3$ together. A similar situation happens for the second system for $x_0$ and $x_2$.

The correct method to address this problem is to break the system of equations into two separate systems of equations. For example, we can split the first system of equations as
\begin{equation}
	\left\{ \begin{matrix}
	2x_1 + 4 x_2 + 2 x_3 = 2 \Delta\\
	\sum_{j=0}^{3} x_j = L-4 \\
	x_j \geq 0 \ \forall j
\end{matrix}	 \right.  \Longrightarrow \left\{ \begin{matrix}
	2x_1 + 4 x_2= 2 \Delta - 2\delta\\
	x_0 + x_1 + x_2 = L-4-\epsilon \\
	2x_3 = 2\delta \\
	x_3 = \epsilon \\
	x_j \geq 0 \ \forall j
\end{matrix}	 \right. \Longrightarrow \left\{ \begin{matrix}
	2x_1 + 4 x_2= 2 \Delta - 2\delta\\
	x_0 + x_1 + x_2 = L-4-\epsilon \\
	0 = 2\delta-2\epsilon \\
	x_3 = \epsilon \\
	x_j \geq 0 \ \forall j
\end{matrix}	 \right.  \ ,
\end{equation}
where we have subtracted the fourth equation from the third one twice. Similarly, for the second one we can write
\begin{equation}
	\left\{ \begin{matrix}
	2x_1 + 2 x_3 = 2 \Delta\\
	\sum_{j=0}^{3} x_j = L-4 \\
	x_j \geq 0 \ \forall j
\end{matrix}	 \right.  \Longrightarrow \left\{ \begin{matrix}
	2x_1= 2 \Delta - 2\delta\\
	x_0 + x_1 = L-4-\epsilon \\
	2x_3 = 2\delta \\
	x_2 + x_3 = \epsilon \\
	x_j \geq 0 \ \forall j
\end{matrix}	 \right.  \ .
\end{equation}
This way, we have split both systems of Diophantine equations into two systems of the equations that have exactly the same form as (\ref{simplifieddiophantine1}). Thus, the number of solutions for a given $\Delta$ and $\delta$ is nothing but the product of the number of solutions of the two systems of equations. However, to do so, we have to pay the price of introducing a sum over all admissible values of $\delta$ and $\epsilon$.

If we use Coeff$\left( P(q) ; q^n \right)$ to denote the coefficient accompanying $q^n$ in the polynomial $P(q)$, we have that the number of solutions for a fixed value of $\Delta$ is
\begin{align*}
	\# \{\text{Solutions of the first system}\}= \sum_{\delta=0}^\Delta \sum_{\epsilon=0}^{L-4} \text{Coeff} \left( \binom{L-2-\epsilon }{2}_q , \Delta - \delta \right) \text{Coeff} \left( \binom{\epsilon }{0}_q , \delta -\epsilon \right) \ , \\
	\# \{\text{Solutions of the second system}\}= \sum_{\delta=0}^\Delta \sum_{\epsilon=0}^{L-4} \text{Coeff} \left( \binom{L-3-\epsilon }{1}_q , \Delta - \delta \right) \text{Coeff} \left( \binom{\epsilon +1 }{1}_q , \delta \right) \ ,
\end{align*}
The second expression has a form that can be immediately interpreted as the coefficient associated to the product of two polynomials, but not the first one. Fortunately, if we take into account the obvious relation Coeff$(P(x),n)=$Coeff$(x^a P(x),n+a)$, the first sum can also be interpreted as the product of two series. Therefore, we can write the following two generating functions
\begin{align}
	F_1 (L,4,2,q) &=\sum q^\Delta \# \{\text{Solutions of the first system}\} = \sum_{\epsilon=0}^{L-4}    q^\epsilon \binom{L-2-\epsilon }{2}_q \binom{\epsilon }{0}_q \ , \notag \\
	F_2 (L,4,2,q) &=\sum q^\Delta \# \{\text{Solutions of the second system}\} = \sum_{\epsilon=0}^{L-4}  \binom{L-3-\epsilon }{1}_q \binom{\epsilon +1 }{1}_q  \ .
\end{align}

However, we have to be careful with the second system of equations due to symmetry. We are interested only on solutions that are inequivalent under translations of the chain. For this case, a solution of the form $(x_0 , x_1 , x_2, x_3)$ is equivalent under translation to a solution of the form $(x_2 , x_3 , x_0 , x_1)$. This means that have to consider only half of the solutions we get. 

Thus, the total number of solutions is given by the function
\begin{equation}
	F (L, 4,2 , q)=L F_1 (L,4,2,q)+ \frac{L}{2} F_2 (L,4,2,q)= \sum_{a=0}^1 \sum_{\epsilon=0}^{L-4} \frac{L}{S_a} q^{(1-a) \epsilon} \binom{L-a-2-\epsilon }{2-a}_q \binom{\epsilon +a }{a}_q \ ,
\end{equation}
where $S_a$ represents a symmetry factor taking care of the overcounting for $a=1$. This means $S_0=1$ and $S_1=2$.

At this point, we can compare our results with those of \cite{Ahn:2021emp} for $L=7$ and $L=8$, as they study these two cases in great detail. We will postpone the comparison for general values of $L$ to the end of this section.

First thing we notice, both our generating functions and the ones computed in \cite{Ahn:2021emp} are the sum of products of two q-binomial coefficient. However, the form of the sum is not the same.

Let us start with $L=7$. The generating function from \cite{Ahn:2021emp} is computed as the sum over the $6$ inequivalent locked states. Our generating function is computed as the sum over the two inequivalent coefficients, $\psi_{34}$ and $\psi_{24}$, and the different forms of splitting the system of linear Diophantine equations. Nevertheless, when evaluated, both sums give the exact same q-binomial coefficients, although dressed with different powers of $q$. In fact, the q-binomial coefficients that the authors of \cite{Ahn:2021emp} find for what they denote as subsectors $1$, $3$, $5$ and $6$ correspond to the q-binomial coefficients that we find for $a=0$ and $\epsilon=0$, $1$, $2$ and $3$ respectively. Similarly, the subsector $2$ corresponds both to $a=1$ and $\epsilon=0$ and $\epsilon=3$, while the subsector $4$ corresponds both to $a=2$ and $\epsilon=1$ and $\epsilon=2$, which shows that the factors $\frac{1}{2}$ is necessary to prevent overcounting.

Similarly, when looking at the $L=8$ case, the two sums are different but they give us equivalent results. In this case, the subsectors $1$, $2$, $5$, $8$ and $6$ map to the cases of $a=0$ with $\epsilon=0,1,\dots ,4$ respectively; subsector $3$ maps to both $a=1$, $\epsilon=0$ and $a=1$, $\epsilon =4$; subsector $4$ maps to both $a=1$, $\epsilon=1$ and $a=1$, $\epsilon =3$; and subsector $7$ maps to $a=1$, $\epsilon=2$. The authors of \cite{Ahn:2021emp} only have to introduce a factor of $\frac{1}{2}$ to subsector $7$ due to symmetry, while we have to introduce the same factor for all the cases with $a=1$. Thus, both sums give rise to exactly the same q-binomial coefficients.

\subsection{The Diophantine equation in the $K=2$ case}

Let us move now to the case of $K=2$ and general $M$. First, we have to count the number of different ways we can arrange the $M-2$ excitations of type $2$ and $2$ excitations of type $3$ that are inequivalent under translations. This is very simple to do: if we fix one of the excitation of type $3$, having the other excitation at a given distance in one direction or the other is exactly equivalent, as we just have to perform a translation that puts the other excitation in the position of the original one. Thus, out of the $M-1$ possible sites where we can put the second excitation of type $3$, we have to consider half of them, giving us a total of $\lceil \frac{M-1}{2} \rceil$ possible inequivalent arrangements, where $\lceil x \rceil$ represents the ceiling function.

Consider then the coefficient $\psi_{M-1 ,M}$, which corresponds to the arrangement $22\dots 233$. In this case, the equation (\ref{Diophgeneral}) takes the form
\begin{displaymath}
\left\{ \begin{matrix}
	\sum_{j=1}^{M} [( n_{M-1} - n_j )+( n_{M} - n_j )]  = n \\
	1\leq n_1 < n_2 <\dots <n_M  \leq L
\end{matrix} \right. \Longrightarrow \left\{ \begin{matrix}
	\sum_{j=1}^{M-2} 2 j x_j + (M-2) x_{M-1} = 2\Delta \\
	\sum_{j=0}^{M-1} x_j = L-M \\
	x_j  \geq 0 \ \forall j
\end{matrix} \right. \ .
\end{displaymath}
Similarly to the $M=4$ case, we can split this system of equations into a system of four linear equations that can be related to q-binomial coefficients
\begin{equation}
\left\{ \begin{matrix}
	\sum_{j=1}^{M-2} 2 j x_j= 2 \Delta - 2\delta\\
	\sum_{j=0}^{M-2} x_j = L-M-\epsilon \\
	(M-2) x_{M-1} = 2\delta \\
	x_{M-1} = \epsilon \\
	x_j \geq 0 \ \forall j
\end{matrix}	 \right.  \ .
\end{equation}
If we subtract the fourth equation from the third $M-2$ times, it becomes $0= 2\delta - (M-2) \epsilon$. Therefore, the number of solutions for a fixed value of $\Delta$ is\footnote{If both $M-2$ and $\epsilon$ are odd, the same reasoning apply after shifting both $\delta$ and $\Delta$ by $\frac{1}{2}$. The same holds for the other configurations.}
\begin{equation}
	\# \{\text{Solutions of the first system}\}= \sum_{\delta=0}^\Delta \sum_{\epsilon=0}^{L-4} \text{Coeff} \left( \binom{L-2-\epsilon }{M-2}_q , \Delta - \delta \right) \text{Coeff} \left( \binom{\epsilon }{0}_q , \delta - \frac{M-2}{2} \epsilon \right) \ .
\end{equation}

For the configuration $22\dots 2323$, the system of equations we find is
\begin{displaymath}
\left\{ \begin{matrix}
	\sum_{j=1}^{M} [( n_{M-2} - n_j )+( n_{M} - n_j )] = n \\
	1\leq n_1 < n_2 <\dots <n_M  \leq L
\end{matrix} \right. \Longrightarrow \left\{ \begin{matrix}
	\sum_{j=1}^{M-3} 2j x_j + (M-4) x_{M-2}+ (M-2) x_{M-1} = 2\Delta \\
	\sum_{j=0}^{M-1} x_j = L-M \\
	x_j  \geq 0 \ \forall j
\end{matrix} \right. \ ,
\end{displaymath}
which can be spit again into a system of four equations
\begin{equation}
\left\{ \begin{matrix}
	\sum_{j=1}^{M-3} 2j x_j= 2 \Delta - 2\delta\\
	\sum_{j=0}^{M-3} x_j = L-M-\epsilon \\
	(M-4) x_{M-2} + (M-2) x_{M-1} = 2\delta \\
	x_{M-2} + x_{M-1} = \epsilon \\
	x_j \geq 0 \ \forall j
\end{matrix}	 \right.  \Longrightarrow  \left\{ \begin{matrix}
	\sum_{j=1}^{M-3} 2j x_j= 2 \Delta - 2\delta\\
	\sum_{j=0}^{M-3} x_j = L-M-\epsilon \\
	2 x_{M-1} = 2 \delta - (M-4) \epsilon \\
	x_{M-2} + x_{M-1} = \epsilon \\
	x_j \geq 0 \ \forall j
\end{matrix}	 \right.\ ,
\end{equation}
where we have subtracted the fourth equation from the third $M-4$ times. With this transformation, both sets of equations have the form we wish, and we can write the number of solutions in terms of products of q-binomial coefficients
\begin{multline}
	\# \{\text{Solutions of the second system}\}=\\
	 \sum_{\delta=0}^\Delta \sum_{\epsilon=0}^{L-4} \text{Coeff} \left( \binom{L-3-\epsilon }{M-3}_q , \Delta - \delta \right) \text{Coeff} \left( \binom{\epsilon +1}{1}_q , \delta - \frac{M-4}{2} \epsilon \right) \ .
\end{multline}

Finally, let us consider the general case, where the two excitations of type $3$ are separated by $a$ excitations of type $2$. In that case, we have
\begin{displaymath}
\left\{ \begin{matrix}
	\sum_{j=1}^{M} [( n_{M-(a+1)} - n_j )+( n_{M} - n_j )] = n \\
	1\leq n_1 < n_2 <\dots <n_M  \leq L
\end{matrix} \right. \Longrightarrow \left\{ \begin{matrix}
	\sum_{j=1}^{M-(a+2)} 2j x_j + \sum_{j=M-(a+1)}^{M-1}(2j-M) x_{j} = 2\Delta \\
	\sum_{j=0}^{M-1} x_j = L-M \\
	x_j  \geq 0 \ \forall j
\end{matrix} \right. \ .
\end{displaymath}
Let us again split the two equations into four
\begin{equation}
\left\{ \begin{matrix}
	\sum_{j=1}^{M-(a+2)} 2j x_j= 2 \Delta - 2\delta\\
	\sum_{j=0}^{M-(a+2)} x_j = L-M-\epsilon \\
	\sum_{j=M-(a+1)}^{M-1} (2j-M) x_{M-j} = 2\delta \\
	\sum_{j=1}^{a+1} x_{M-j} = \epsilon \\
	x_j \geq 0 \ \forall j
\end{matrix}	 \right.  \Longrightarrow  \left\{ \begin{matrix}
	\sum_{j=1}^{M-(a+2)} 2j x_j= 2 \Delta - 2\delta\\
	\sum_{j=0}^{M-(a+2)} x_j = L-M-\epsilon \\
	\sum_{j=1}^{a-1} 2j x_{M-j} = 2\delta -(M-2a - 2) \epsilon\\
	\sum_{j=1}^{a+1} x_{M-j} = \epsilon \\
	x_j \geq 0 \ \forall j
\end{matrix}	 \right.\ ,
\end{equation}
where we have subtracted the fourth equation from the third $M-2a-2$ times. After this transformation, both sets of equations have the correct form, and we can write the number of solutions in terms of products of q-binomial coefficients
\begin{multline}
	\# \{\text{Solutions of the (a+1)-th system}\} = \\
	\sum_{\delta=0}^\Delta \sum_{\epsilon=0}^{L-4} \text{Coeff} \left( \binom{L-2-a-\epsilon }{M-2-a}_q , \Delta - \delta \right) \text{Coeff} \left( \binom{\epsilon +a}{a}_q , \delta - \frac{M-2a - 2}{2} \epsilon \right) \ .
\end{multline}
Therefore, for any separation $a$, the number of solutions is always given by a product of two q-binomial coefficients.

Before putting everything together, we have to make a comment on the symmetry under translations of our solutions. In the case of $M=4$, we showed that the arrangement $2323$ overcounts the number of solutions. This issue arises for all even values of $M$ when $a=\frac{M-2}{2}$, because a solution $(x_0, \dots ,x_{M-1})$ is equivalent under translations to the solution $(x_{M/2} , \dots , x_{M-1} , x_0 , \dots ,x_{(M-2)/2})$. Consequently, we have to count only half of the solutions for this particular value of $a$.

Now that we have all the necessary pieces, we can write the generating function of the number of solutions
\begin{equation}
	F(L,M,2,q)=\sum_{a=0}^{\lceil \frac{M-1}{2} \rceil} \sum_{\epsilon=0}^{L-M} \frac{L}{S_a} q^{\frac{M-2a - 2}{2} \epsilon} \binom{L-2-a-\epsilon }{M-2-a}_q \binom{\epsilon +a}{a}_q \ ,
\end{equation}
where the symmetry factor $S_a=1+\delta_{a, (M-2)/2}$, that is, is equal to $1$ for every $a$ except when $M$ is even and $a=\frac{M-2}{2}$, where it is equal to $2$.

\subsection{The Diophantine equation for general $K$}

Now that we have understood the case of $K=2$, we can move to the most general case.

First and foremost, we have to calculate the number of arrangements of $M-K$ excitations of type $2$ and $K$ excitations of type $3$ that are inequivalent under translations. This is equivalent to the problem of counting the number of necklaces with $M-K$ beads of one type and $K$ beads of another type. The solution to this counting problem is known to be \cite{A037306}
\begin{equation}
	\frac{1}{M} \sum_d \phi(d) \binom{M/d}{K/d} 
\end{equation}
where the sum is taken over the divisors of the greatest common divisor of $M$ and $K$, and $\phi (d)$ is the Euler totient function.

Regarding the Diophantine equations we need to solve, the process is the same as for $K=2$ mutatis mutandis. First, we change variables to relative positions; after this, we split the two equations into $2K$ equations that have the same form as (\ref{simplifieddiophantine1}); thanks to this splitting, we can check that the generating function of the number of solutions for a given arrangement of $2$'s and $3$'s can be written as the product of $K$ q-binomial coefficients; finally, we have to sum all these generating functions, multiplied by a factor of $L$ to take into account all the possible values of the total momentum, and divided by a possible symmetry factor.

To exemplify the first three steps, we will explicitly compute the generating function for $K=3$ with separations $a_1+1$ and $a_2+1$. This means that the last $3$ and the next-to-last $3$ are separated by $a_1$ 2's, while the next-to-last and the third-to-last are separated by $a_2$ $2$'s. In terms of the positions of the excitations, the Diophantine equation reads
\begin{displaymath}
\left\{ \begin{matrix}
	\sum_{j=1}^{M} [( n_{M-(a_1+a_2+2)} - n_j )+( n_{M-(a_1+1)} - n_j )+( n_{M} - n_j )] = n \\
	1\leq n_1 < n_2 <\dots <n_M  \leq L
\end{matrix} \right. \ . 
\end{displaymath}
which transform into the following system of linear Diophantine equations in terms of the relative positions
\begin{displaymath}
	\left\{ \begin{matrix}
	\sum_{j=1}^{M-(a_1+a_2+3)} 3j x_j + \sum_{j=M-(a_1+a_2+2)}^{M-(a_1+2)}(3j-M) x_{j} +\sum_{j=M-(a_1+1)}^{M-1}(3j-2M) x_{j}= 3\Delta \\
	\sum_{j=0}^{M-1} x_j = L-M \\
	x_j  \geq 0 \ \forall j
	\end{matrix} \right. \ . 
\end{displaymath}
We can now spit these two equations into six
\begin{equation}
	\left\{ \begin{matrix}
	\sum_{j=1}^{M-(a_1+a_2+3)} 3j x_j= 3\Delta-3\delta_1 - 3\delta_2 \\
	\sum_{j=0}^{M-(a_1+a_2+3)} x_j = L-M-\epsilon_1-\epsilon_2 \\
	\sum_{j=M-(a_1+a_2+2)}^{M-(a_1+2)}(3j-M) x_{j}= 3\delta_2 \\
	\sum_{j=M-(a_1+a_2+2)}^{M-(a_1+2)} x_j = \epsilon_2 \\
	\sum_{j=M-{a_1+1}}^{M-1} (3j-2M) x_{j}= 3\delta_1 \\
	\sum_{j=M-{a_1+1}^{M-1}} x_j = \epsilon_1 \\
	x_j  \geq 0 \ \forall j
	\end{matrix} \right. \Longrightarrow  \left\{ \begin{matrix}
	\sum_{j=1}^{M-(a_1+a_2+3)} 3j x_j= 3\Delta-3\delta_1 - 3\delta_2 \\
	\sum_{j=0}^{M-(a_1+a_2+3)} x_j = L-M-\epsilon_1-\epsilon_2 \\
	\sum_{j=0}^{a_2}3j x_{M-(a_1+a_2+2)+j}= 3\delta_2- [2M-3(a_1+a_2+2)] \epsilon_2 \\
	\sum_{j=0}^{a_2} x_{M-(a_1+a_2+2)+j} = \epsilon_2 \\
	\sum_{j=0}^{a_1} 3j x_{M-(a_1+1)+j}= 3\delta_1 - [M-3a_1] \epsilon_1 \\
	\sum_{j=0}^{a_1} x_{M-(a_1+1)+j} = \epsilon_1 \\
	x_j  \geq 0 \ \forall j
	\end{matrix} \right. \ , 
\end{equation}
where have performed the appropriate subtractions of equations. As these equations now have the correct form to be interpreted as q-binomial coefficients, we can write
\begin{equation}
	F_{a_1 , a_2} (M,L,3,q)=\sum_{\epsilon_1=0}^{L-M} \sum_{\epsilon_2=0}^{L-M-\epsilon_1} \frac{L  q^{\kappa}}{S_{a_1 ,a_2}} \binom{L-(a_1+a_2+3)-\epsilon_1-\epsilon_2 }{M-(a_1+a_2+3)}_q \binom{\epsilon_2 +a_2 }{a_2}_q \binom{\epsilon_1+a_1 }{a_1}_q \ ,
\end{equation}
where $\kappa= \frac{[2M-3(a_1+a_2+2)]\epsilon_2 + [M-3a_1] \epsilon_1}{3}$ and $S_{a_1 ,a_2}$ is a symmetry factor that counteracts a possible overcounting from configurations that are equivalent under translations.

As the construction is similar for any value of $K$, we can write the following general expression
\begin{equation}
	F(M,L,K,q)=\sum_{\{a\}} \sum_{\epsilon_1=0}^{L-M} \sum_{\epsilon_2=0}^{L-M-\epsilon_1} \dots \sum_{\epsilon_{K-1}=0}^{L-M-\epsilon_1-\dots -\epsilon_{K-2}} \frac{L  q^{\kappa}}{S_{\{a\}}}  \binom{L+K+\sum_{j=1}^{K-1} (a_j-\epsilon_j)}{M-K-\sum_{j=1}^{K-1} a_j}_q \prod_{j=1}^{K-1} \binom{\epsilon_j +a_j }{a_j}_q \ ,
\end{equation}
where $\kappa=\frac{\sum_{j=1}^{K-1} [j M - K \sum_{k=1}^j (a_k+1)] \epsilon_j }{K}$, and the sum $\sum_{\{a\}}$ is taken over all the arrangements of $2$'s and $3$'s that are not equivalent under translations.

The symmetry factor $S_{\{a\}}$ is usually not easy to compute, as it counteracts overcountings due to some configurations having additional symmetry under translations. Nevertheless, these configurations can only arise if and only if $M$ and $K$ have common prime factors, as otherwise we cannot arrange the excitations in a configuration with enhanced symmetry. Thus, $S_{\{a\}}=1$ for all configurations, provided $M$ and $K$ are coprime.

Finally, now that we have an expression valid for all possible values of $L$, $M$ and $K$, we can compare our result with equation (4.30) of \cite{Ahn:2021emp}.

Similarly to the $M=4$ and $K=2$ case we discussed above, both generating functions are the sum of products of $K$ q-binomial coefficients, but the rules to compute which terms enter the sum are completely different. In our approach, we have to sum over the all possible coefficients of the wavefunction (\ref{wavefunctiongeneralMK}) that are inequivalent under translations (and have the excitations of type $3$ as close together as possible and as close to the right as possible), and then sum over how to split the system of linear Diophantine equations. In the approach of \cite{Ahn:2021emp}, they sum over all the possible ways of arranging the excitations of type $1$ and type $2$ around the $K$ excitations of type $3$ that give rise to states that can be written as tensor products of locked states, and use that each of these states are in a one-to-one correspondence with a Jordan chain, as they are always the generalised eigenvector of rank one of one of these Jordan chains (although this identification can be spoiled by chain mixing). From that observation and the definition of generalised eigenvector, they are able to show that the length of each Jordan chain can be computed as this product of q-binomial coefficients.

Although not obvious, these two approaches are actually equivalent. Our sum over coefficients that are inequivalent under translations is equivalent to their sum over all the possible ways of arranging the excitations of type $2$, while the sum over $\epsilon_i$'s that we get from splitting the system of equations is equivalent to their sum over all the possible ways of arranging the excitations of type $1$. The only difference is how we treat the cases when $S_{\{a\}} \neq 1$. In our computation, we count all the possible ways of arranging the excitations of type $1$, without considering that sometimes we are counting arrangements that are equivalent under translations, as the symmetry factor will take care of it. In contrast, the authors of \cite{Ahn:2021emp} already take this overcounting into consideration in their sum, so they only consider inequivalent arrangements and use a smaller or no symmetry factors. Nevertheless, both methods give the same result at the end.

The fact that our expression overcounts all the configurations with $S_{\{a\}} \neq 1$ instead of carefully avoiding them works in our favour. Thanks to it, the most difficult part of the sum to evaluate, which is the sum over inequivalent configurations, is independent of the length of the chain. The price to pay are the sums over $\epsilon_i$'s, but these are regular sums, making them easier to compute compared to the ones appearing in \cite{Ahn:2021emp}.

\section{Conclusions}

In this article, we have finished the characterisation of the Jordan chains associated to the eclectic  three-state spin chain introduced in \cite{Ipsen:2018fmu} that we started in \cite{firstpart}.

In our previous article, we described a method to construct the generalised eigenvectors of a defective matrix that is obtained as the exceptional point of a diagonalisable matrix by considering limits of linear combinations of the eigenvectors of the latter. After constructing the eigenvectors of the spin chain at finite values of the twist for $K=1$ using a modified version of the Nested Coordinate Bethe Ansatz, we applied that recipe to construct the generalised eigenvectors of the eclectic spin chain, and we showed that the information about the length of the Jordan chains for this subsector can be encoded into a q-deformed binomial coefficient. In this article, we have used the same procedure to show that the length of Jordan chains for general values of $K$ can be encoded into a generating function that is constructed out of sums of products of $K$ of those q-deformed binomial coefficients.

The counting of states of this eclectic spin chain was started in \cite{StaudacherAhn}, where the authors described the coalescence of eigenvectors, which is a clear sign that the strongly twisted limit they studied corresponds to an exceptional point of the Hamiltonian. This counting, which was incomplete, was refined in \cite{Ahn:2021emp} using the fact that the tensor product of locked states are always the generalised eigenvector of the lowest rank of a Jordan chain, together with combinatorial arguments and the definition of generalised eigenstate. This allowed them to codify the structure of Jordan chains in terms of sums of products of q-deformed binomial coefficients.

The method used in this article to classify the Jordan chains and the one used in \cite{Ahn:2021emp} are completely different, so an agreement between both results would only strengthen their validity. Despite both being sums of products of Gaussian polynomials, the addends are equal but the sums are not the same. Thankfully, this mismatch is only apparent, and both sums end up yielding the same result. 

However, neither these articles nor ours completely finish the task of classifying the Jordan chains of this eclectic spin chain. All these results rest on the unproven conjecture that chain mixing does not spoil them. Currently, it has only been proven that the longest Jordan chains remain intact unless $\xi_i + \xi_j \neq 0$, for any two deformation parameters. A method to discern if our naïve identification of Jordan chains is correct was proposed in \cite{firstpart}, which uses properties of the left eigenvectors to go up and down a Jordan chain, but it becomes cumbersome to apply really fast. Thus, it remains to check for which values of the deformation parameters $\xi_i$ the conjecture regarding chain mixing is correct.

Another interesting line of work would be to apply the method used here to other non-diagonalisable $R$-matrices. All the integrable systems with nearest-neighbour interaction and either $\mathfrak{su}(2)$ or $\mathfrak{su}(2) \oplus \mathfrak{su}(2)$ symmetry were classified in \cite{DeLeeuw:2019gxe} and \cite{DeLeeuw:2019fdv,deLeeuw:2020ahe,deLeeuw:2020xrw} respectively. Some of the $R$-matrices that appear in these classifications are non-diagonalisable, making their transfer matrices good candidates for the algorithm presented here.

\section{Acknowledgements}

We are very thankful to Alessandro Torrielli for helpful discussions. We are grateful to Alessandro Torrielli, Leander Wyss and Luke Corcoran for reading the manuscript and providing very useful comments. This work is supported by the EPSRC-SFI grant EP/S020888/1 {\it Solving Spins and Strings}. 

No data beyond those presented and cited in this work are needed to validate this study.



\end{document}